\documentclass[fp,twocolumn]{jpsj3}
\usepackage{txfonts}
 \usepackage{xcolor}

\title{Pulsed quantum annealing}

\author{Vasilios Karanikolas\thanks{Present address: International Center for Young Scientists (ICYS), National Institute
for Materials Science (NIMS) 1-1 Namiki, Tsukuba, Ibaraki 305-0044,
Japan. E-mail:KARANIKOLAS.Vasileios@nims.go.jp} and Shiro Kawabata\thanks{Corresponding author. E-mail:s-kawabata@aist.go.jp}}
\inst{Device Technology Research Institute, National Institute of Advanced
Industrial Science and Technology (AIST), Tsukuba, Ibaraki 305-8568,
Japan}

\abst{We propose a modified quantum annealing protocol, $i. e.,$ {\it pulsed quantum annealing} (PQA), in order to increase the success probability by a pulse application during the quantum annealing process.
It is well known that the success probability of the conventional quantum annealing is reduced due to the Landau-Zener transitions. 
By applying a pulse to the system, we modulate the success probability and increase it, compared to the 
conventional quantum annealing, by optimizing the pulse parameters. 
We demonstrate our findings for a single qubit both numerically and analytically. The analytical model is based on the tranfer matrix approach and it is in good agreement with the full numerical results. 
We also investigate the PQA protocol for multi-qubit cases $i. e.,$ random spin-glass instances, and we present an overall increase of the success probability over the conventional quantum annealing, by optimizing the pulse parameters. 
Our results indicate that PQA can be used to design future high-performance quantum annealing machines, especially for hard instances that the conventional QA protocol behaves poorly.}

\begin{document}
\maketitle

\section{Introduction}

Quantum annealing (QA) has been introduced as an alternative to
simulated annealing for efficiently solving combinatorial optimization
problems \cite{Kadowaki1998,Farhi2000}. The advantage of QA
over the  classical simulated annealing \cite{Kirkpatrick1983} is that during
the annealing, the search for the optimum state of the system can stuck to a local
minima, then quantum fluctuations can help the system to tunnel from
the local minima to the global one \cite{Das2008}. The research on
QA is intensified in recent years, after the commercialization of
the QA machines by D-Wave Systems Inc. using superconducting flux qubits
\cite{Harris2010,Johnson2011,Dickson2013}. The QA machines from D-Wave
Inc. have been used in a number of diverse hard optimization problems,
to name a few: global warming \cite{Raymond2016}, traffic control
\cite{Neukart2017}, election forecasting\cite{Henderson2019}, online advertisement allocation\cite{Tanahashi2019}, and analyzing data regarding the Higgs boson discovery
from Large Hadron Collider \cite{Mott2017}. In addition, new 
quantum annealing machines are now developed by several groups \cite{Barends2016,Maezawa2017,Weber2017,Novikov2018,Mukai2019,Maezawa2019,Tanamoto2019,Goto2019}.

The adiabatic theorem guaranties that if the annealing process is slow enough
the final state will be the ground state of the problem Hamiltonian,
thus a solution of an optimization problem. The annealing time is
inversely proportional to energy gap between the ground and the first
excited state \cite{Albash2016}. Hence, optimization problems with
extremely small energy gap demand long annealing times.
However, long annealing times  reduces the success probability (SP) due to coupling 
of the qubits with the environment and decoherence \cite{Albash2015}. As SP is defined 
the probability of the instantaneous ground state of the Hamiltonian at the end of the annealing. Thus, SP measures how close
we are to the optimum solution. On the same time, for spin
glass problems it has been found that decreasing the annealing time can be beneficial \cite{Steiger2015,Crosson2014}.
The physical explanation is that for fixed annealing times the state
can stuck in a local minima, and by reducing the annealing time, non-adiabatic
processes are introduced that can kick the system from the local minima of the system, to the global minimum. Thus, an induced nonadiabatic process can be used to increase the SP.

Importantly, the QA is closely connected with the Landau-Zener physics\cite{Shevchenko2010}. 
The transition probability at avoided energy level crossings between the ground and first excited state has been investigated intensively \cite{Zener1932}. Vitanov $et$ $al.,$ have showed that a diabatic pulsed quantum fluctuation during the Landau-Zener transition causes an oscillatory behavior of the transition probability\cite{Vitanov1996}.

In this paper we introduce the {\it pulsed quantum annealing} (PQA) protocol.
By applying a single pulse to the system during the quantum annealing process, we can modulate the SP and increase it in comparison with the conventional quantum annealing, for a set of pulse parameters.
Starting from the single qubit case, we analytically show that this modulation can be explained by the destructive and constructive interference model based on the transfer matrix approach\cite{Ashhab2007,Zhou2014,Silveri2015}. To our knowledge, that is the first time that this semianalytical model is used to describe the quantum annealing protocol.
In addition, we also numerically investigate PQA for multi-qubit cases and confirm the enhancement of SP by optimizing pulse parameters over multiple random instances.

The paper is organized as follow. 
In Sec.\ref{sec:II} we consider
the PQA protocol for a single qubit. 
We present the transfer matrix method\cite{Ashhab2007},
using the sudden approximation, and compare it with full numerical
results. 
In Sec.\ref{sec:III} we apply PQA for multi-qubit cases.  
We show that for any instance
generated, for certain pulse parameters, the SP is increased,
compared to conventional QA. 
In Sec.\ref{sec:IV} we give the concluding
remarks and discuss possible future directions. In Appendix A
\ref{sec:Appendix-A} we present in more details the transfer matrix
method using the sudden approximation.

\section{Pulsed quantum annealing for a single qubit\label{sec:II}}

\subsection{Transfer-matrix approach to describe the pulsed quantum annealing for a single qubit}

The Hamiltonian describing the PQA protocol for a single qubit, including the applied pulse,
has the form,
\begin{equation}
H(t)=\left[ \frac{t}{t_{f}} \varepsilon+C\Lambda_{P}(t)\right]\sigma_{z}+\left[1-\frac{t}{t_{f}}\right]\Delta\sigma_{x},\label{eq:01}
\end{equation}
where $\sigma_{z}$ and $\sigma_{x}$ are the Pauli matrices, $t_{f}$
is the annealing time, $\Delta$ gives the strength of the quantum
fluctuations, $\varepsilon$ is the energy difference between the
$\left|0\right\rangle $ and $\left|1\right\rangle $ states in the
diagonal term and $C$ is the strength of the applied diabatic pulse.
$\Lambda(t)=\Theta(t-t_{C}+t_{D}/2)\Theta(t_{C}+t_{D}/2-t)$ is the
pulse shape, in the diagonal part, which is characterized by the
pulse center $t_{C}$ and the pulse duration $t_{D}$; where the theta function is defined as 
$\Theta(x)=1$ if $x>0$. Through out
this paper we use energy units expressed through $\Delta$, thus the
time scales are expressed in $\hbar/\Delta$ units. When there is
no applied pulse we have the conventional linear ramp QA, which is
characterized by the usual Landau-Zener physics at the avoided crossing.
Throughout this paper we focus on cases where the temperature is
well below the minimum energy gap, $E_{G}^{min}\gg k_{B}T$, thus
we take the zero temperature limit, $T=0K$.

Our approach consists of solving numerically the time-dependent
Schr\"{o}dinger equation
\begin{equation}
i\hbar\frac{\partial}{\partial t}\left|\psi(t)\right\rangle =H(t)\left|\psi(t)\right\rangle ,\label{eq:02}
\end{equation}
which is used to simulate the QA process, with and without the diabatic
pulse. Also, the exact diagonalization of the Hamiltonian $H(t)$~\eqref{eq:01},
without the applied pulse, is used to calculate its instantaneous
eigenstates $\left|\psi_{i}(t)\right\rangle $, for $i=0,1,\ldots,2^{n}-1$,
for $n$ qubit systems. At the end of the annealing, $t=t_{f}$, we
define the SP as the projection of the instantaneous ground state
$\left|\psi_{0}(t_{f})\right\rangle $ to the system state $\left|\psi(t_{f})\right\rangle $
\begin{equation}
P=\left|\left\langle \psi_{0}(t_{f})\left|\psi(t_{f})\right.\right\rangle \right|^{2}.\label{eq:03}
\end{equation}

The conventional adiabatic QA protocol has to fulfill the adiabatic
condition \cite{Albash2016}
\begin{equation}
t_{f}\gg\frac{max_{0\leq s\leq1}\left[\left\langle \psi_{0}(s)\right|\frac{dH(s)}{ds}\left|\psi_{1}(s)\right\rangle \right]}{min_{0\leq s\leq1}\left[E_{01}(s)\right]^{2}},\label{eq:04}
\end{equation}
for completely reducing unwanted Landau-Zener transitions, where $s=t/t_{f}$,
$\left|\psi_{i}(s)\right\rangle $ describe the ground, $i=0$, and
first, $i=1$, excited instantaneous eigenstates of the Hamiltonian
Eq.~\eqref{eq:01}, without the pulse term, and $E_{01}(s)$ the
energy difference between them. The physical meaning of Eq.~\eqref{eq:04}
is that the annealing process should be long enough for suppressing undesired Landau-Zener transitions.

\begin{figure*}[t]
\includegraphics[width=0.4\textwidth]{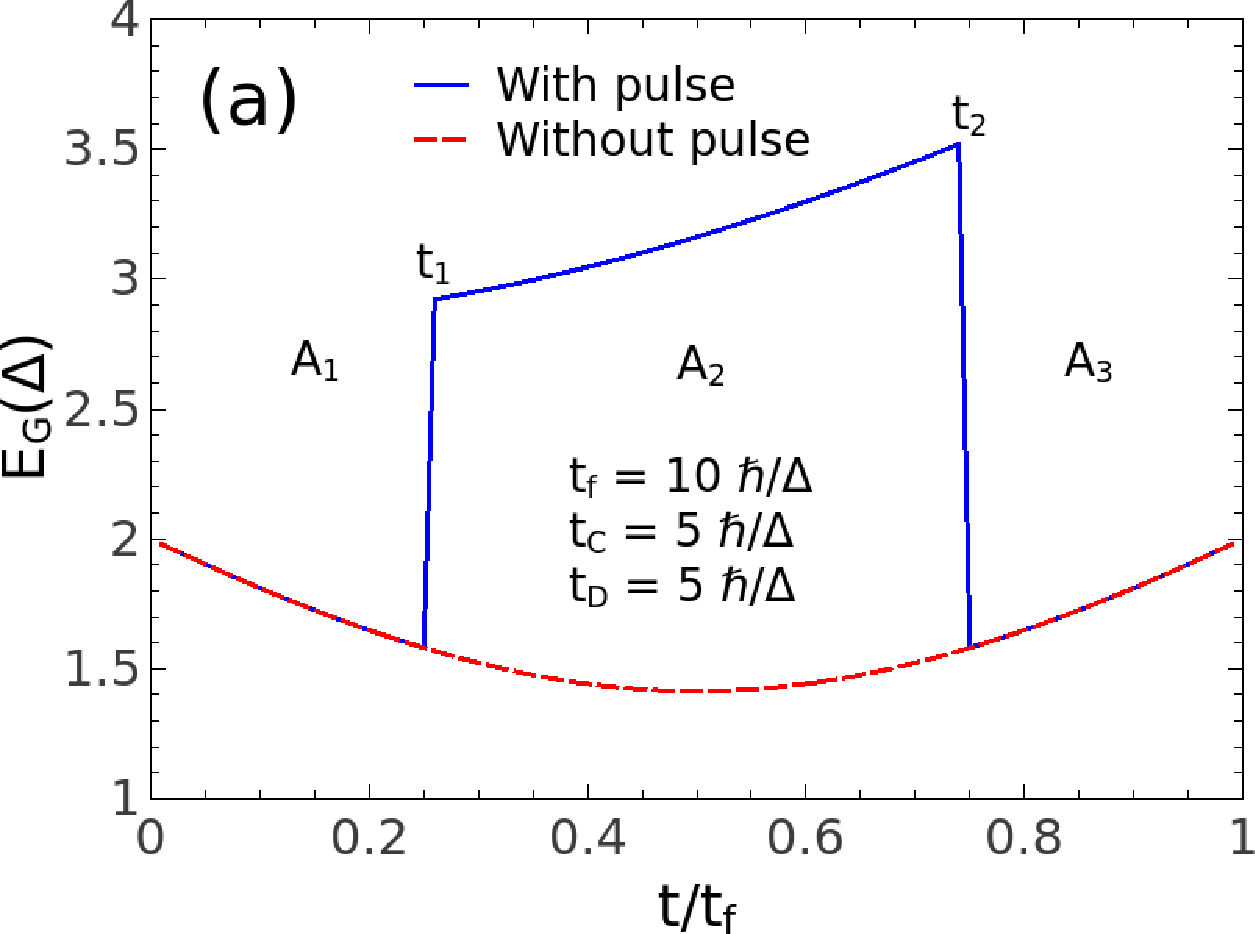}~~~\includegraphics[width=0.4\textwidth]{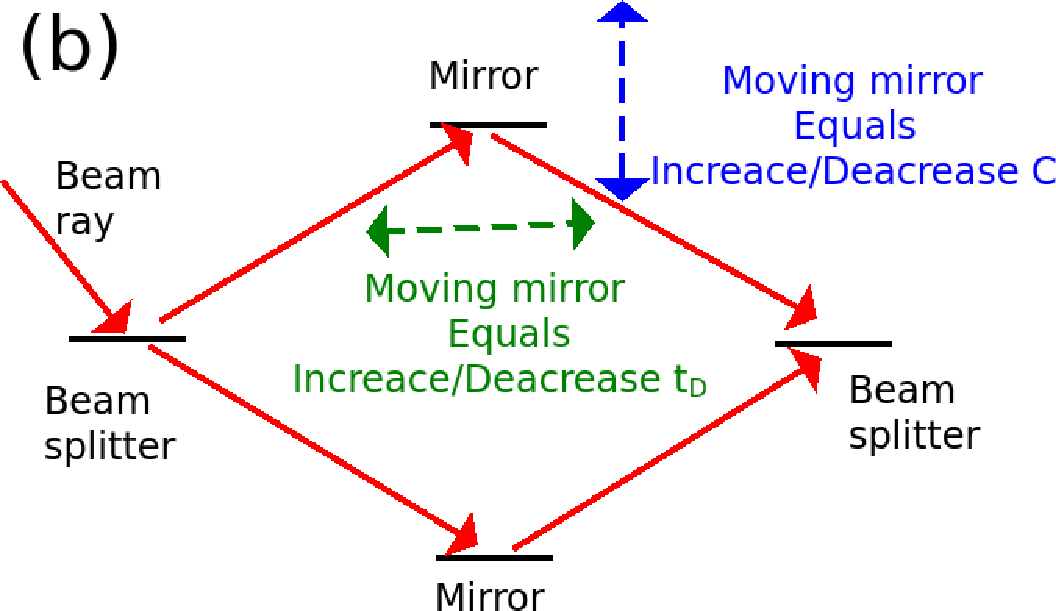}
\caption{
(Color online) (a) The energy gap, $E_{G}(t)$, between the ground
and the excited states for a single qubit by varying the time $t$
during the annealing time $t_{f}=10\hbar/\Delta$ and $\varepsilon=1\Delta$.
The blue continuous line gives the $E_{G}(t)$ in the presence of
the pulse, with $t_{C}=5\hbar/\Delta$, $t_{D}=5\hbar/\Delta$ and
$C=1\Delta$, and the red dashed line for the $C=0$ case of conventional
QA. (b) The Mach-Zehnder interferometer and its analogy to the modulation
of the success probability by a diabatic pulse application to a single
qubit, of strength $C$ and duration $t_{D}$.
}
\label{fig:Fig01}
\end{figure*}

The energy difference between the ground and the excited states of
Eq.~\eqref{eq:01} is given by $E_{G}(t)=2\sqrt{\left(1-t/t_{f}\right)^{2}
\Delta^{2}+\left(t/t_{f}\varepsilon+C\Lambda_{P}(t)\right)^{2}}$
and is plotted in Fig.~\ref{fig:Fig01} for the case of $t_{f}=10\hbar/\Delta$
and $\varepsilon=1\Delta$ with the pulse, continuous line, and without
the pulse, dashed line. The pulse parameters are $t_{C}=5\hbar/\Delta$,
$t_{D}=5\hbar/\Delta$ and $C=1\Delta$. We observe that the pulse
application has split the time evolution path of the qubit's states
in three parts, denoted by the times $t_{1,2}=t_{C} \pm t_{D}/2$,
defining the pulse application. 

The pulse
application is a fast diabatic process, which violates the adiabatic
condition at the pulse application times $t_{1}$ and $t_{2}$. Away
from $t_{1}$ and $t_{2}$, in the regions $j=A_1,A_2,A_3$, the qubit follows
essentially an adiabatic evolution path as long as Eq.~\eqref{eq:04}
is satisfied. Then, the time dependent state
of the single qubit, described by the state vector  $\mathbf{b}(t)=\left(\begin{array}{c}
b_{0}(t)\\b_{1}(t)\end{array}\right)$
(see Appendix A), during these regions can be
found by $\mathbf{b}(t_{q})=U_{j}(t_{q},t_{s})\mathbf{b}(t_{s})$,
where $j=A_1,A_2,A_3$. The unitary evolution matrix is defined as 
\begin{equation}
U_{j}(t_{q},t_{s})=\left(\begin{array}{cc}
e^{-i\zeta_{j}(t_{q},t_{s})} & 0\\
0 & e^{i\zeta_{j}(t_{q},t_{s})}
\end{array}\right)=e^{-i\zeta_{j}(t_{q},t_{s})\sigma_{z}},\label{eq:05}
\end{equation}
where
\begin{equation}
\zeta_{j}(t_{q},t_{s})=\int_{t_{s}}^{t_{q}}E_{G}(t)dt,\label{eq:06}
\end{equation}
is the phase acquired during the adiabatic evolution in regions $j=A_1,A_2,A_3$.
The acquired phases $\zeta_{j}(t_{q},t_{s})$ have a geometrical interpretation,
they are connected with the area bellow the curve in Fig.~\ref{fig:Fig01}.
$t_{s}$ and $t_{q}$ define the start and finish times at each region
$A_1,A_2$ and $A_3$.

At times $t_{1}$ and $t_{2}$ the Hamiltonian changes rapidly due
to the pulse application; the change is so sudden that the system
does not have time to readjust its state \cite{Messiah2014}, thus the
qubit state at $t_{l}^{+}$ can be written in terms of the state at
$t_{l}^{-}$, for $l=1,2$, where $t_{l}^{\pm}$ are the times right
after and right before the transition times imposed by the pulse.
At these times the qubit state
is described by $\mathbf{b}(t_{l}^{+})=N_{l}\mathbf{b}(t_{l}^{-})$,
where the matrices $N_{l}$ have the form \cite{Silveri2015}

\begin{equation}
N_{l}=\left(\begin{array}{cc}
\sqrt{1-p_{s}} & \sqrt{p_{s}}\\
\sqrt{p_{s}} & \sqrt{1-p_{s}}
\end{array}\right),\label{eq:07}
\end{equation}
where $\sqrt{1-p_{s}}=\left\langle \psi_{0}^{+}|\psi_{C}^{-}\right\rangle $
and $\sqrt{p_{s}}=\left\langle \psi_{0}^{-}|\psi_{C}^{-}\right\rangle $,
the states $\left|\psi_{0,C}^{\pm}\right\rangle $ are the instantaneous
eigenstates of the Eq.~\eqref{eq:01} without, $0$, and with, $C$,
the pulse application, more details can be found in Appendix A.
The usual Landau-Zener transitions do not
characterize the transition at the times $t_{1}$ and $t_{2}$, where
the pulse is applied. The pulse application discussed in this
paper is an inherently diabatic process. 

Concretely, the total evolution of the qubit state, starting at time
$0$, at the end of the annealing time, $t_{f}$, is described by,
\begin{equation}
\left|\psi(t_{f})\right\rangle =U(t_{f},t_{2})N_{2}U(t_{2},t_{1})N_{1}U(t_{1},0)\left|\psi(0)\right\rangle \label{eq:08}
\end{equation}
The transfer matrix model described in this paper is inspired by the adiabatic-impulse model \cite{Garraway1997,Damski2005,Damski2006}, describing the evolution of the quantum system considered, except
at the times $t_{1}$ and $t_{2}$ where is described in the sudden approximation
regime. The sudden approximation has been used to explain the Stuckelberg interference in superconducting qubits caused by periodic latching modulation \cite{Silveri2015}. To our knowledge, it is the first time such physical explanation applied in the context of a QA protocol.

\subsection{Tuning the success probability using the pulsed quantum annealing protocol for a single qubit}

\begin{figure*}
\includegraphics[width=0.4\textwidth]{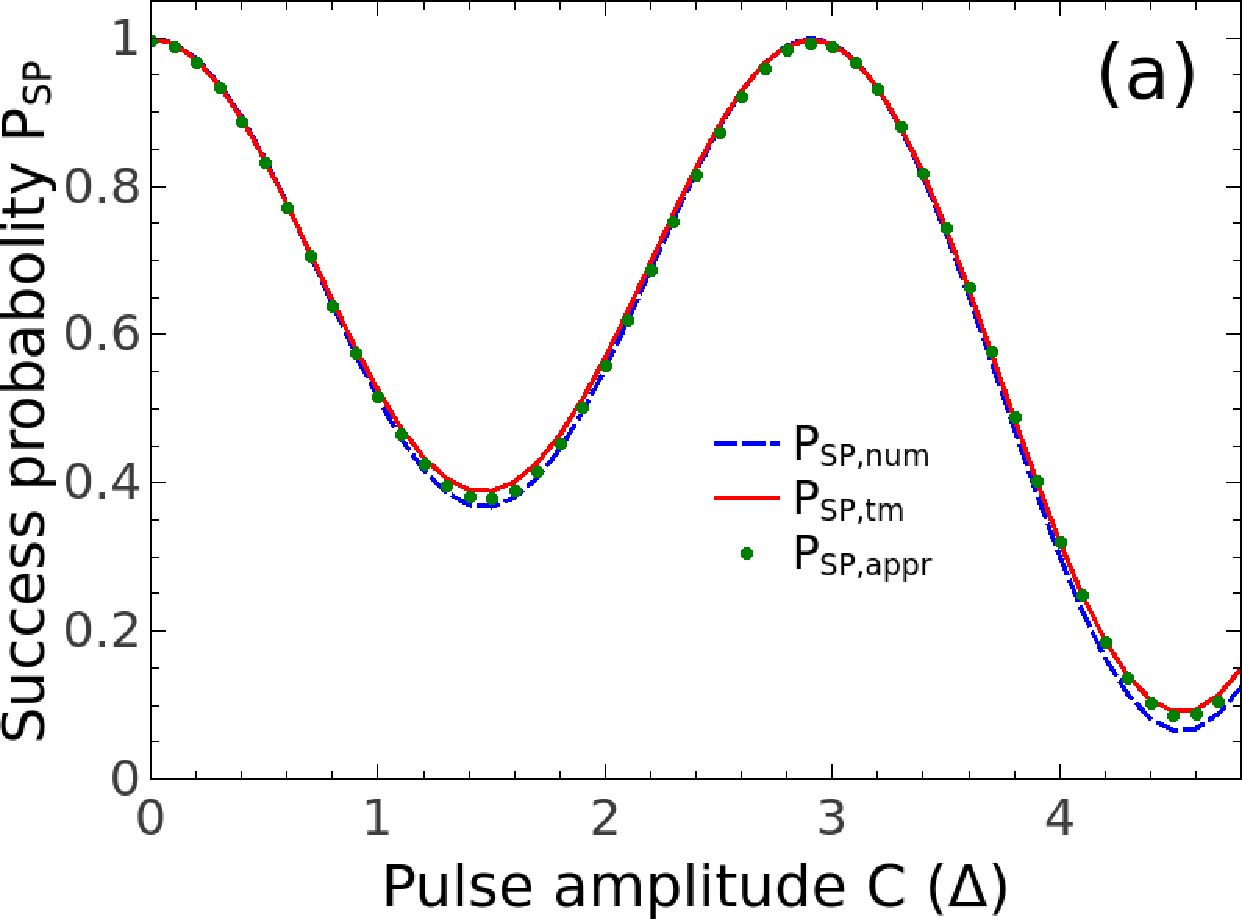}~~~~~\includegraphics[width=0.4\textwidth]{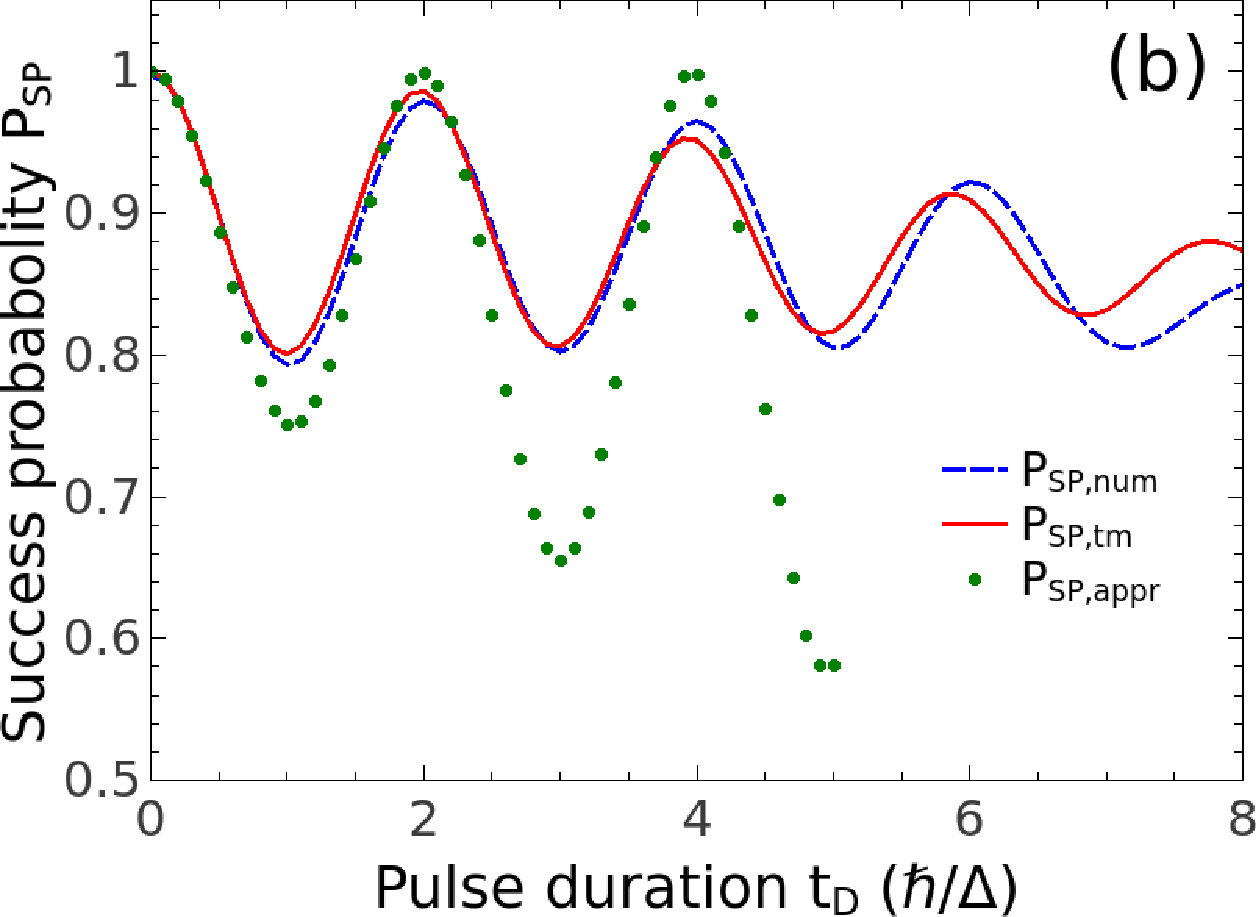}
\caption{(Color online) We present the success probability (SP), the probability of being occupied the ground state, of a
qubit  by varying the pulse parameters $C$ and $t_{D}$,
when the annealing time is $t_{f}=10\hbar/\Delta$ and $\varepsilon=1\Delta$,
for which the adiabatic condition holds for $C=0$. (a) The pulse
amplitude $C$ is varied for fixed pulse center $t_{C}=1\hbar/\Delta$
and pulse duration $t_{D}=1\hbar/\Delta$, (b) the pulse duration
$t_{D}$ is varied for fixed value of the pulse center $t_{C}=5\hbar/\Delta$
and pulse amplitude $C=1\Delta$. The blue dashed line gives the full
numerical results, $P_{SP,num}$, the red continuous line the transfer
matrix approach, $P_{SP,tm}$, and the green dots the approximate
expression $P_{SP,appr}$.}
\label{fig:02}
\end{figure*}

We start by investigating the PQA protocol for a single qubit, we present
results from the full numerical and analytical methods introduced in the previous
subsection. In Fig.~\ref{fig:02} we present the SP, $P_{SP}$, for an annealing
time $t_{f}=10\hbar/\Delta$, for which the adiabatic condition for the conventional QA ($C=0$)
holds, and fixed $\varepsilon=1\Delta$.
In Fig.~\ref{fig:02}(a) we vary the pulse amplitude, $C$,
for a fixed value of the pulse center, $t_{C}=1\hbar/\Delta$, and of the pulse duration,
$t_{P}=1\hbar/\Delta$, while in Fig.~\ref{fig:02}(b) we vary the
pulse duration, $t_{P}$, for fixed value of the pulse center, $t_{C}=5\hbar/\Delta$,
and of the pulse amplitude, $C=1\Delta$. The blue dots give the SP for
the case the Schr\"{o}dinger equation is solved numerically, using the
Hamiltonian of Eq.~\eqref{eq:01}, and the red continuous line gives
the results using the transfer matrix approach described earlier.
We observe a good agreement between the two methods. The observed
oscillations in the SP, $P_{SP}$, are due to destructive and constructive
interference caused by the phase accumulation the qubit acquires during
the pulse application. In Fig.~\ref{fig:Fig01}(b) a direct analogy
can be drown with a Mach-Zehnder interferometer \cite{Oliver2005}
which is composed from two beam splitters. The first beam splitter divides the
optical beam into two coherent beams that can follow different paths.
The second beam splitter recombines and superimpose these beams, the
different paths followed give interference fringes. To sum up, the
pulse application due to interference effect, caused by the pulse
application, can increase the SP for varying the pulse parameters.
If we consider the case that $\zeta_{1}(0,t_{1})\simeq\zeta_{3}(t_{2},t_{f})$
then the SP can be approximated by the expression $P_{SP,appr}=1-4p_{s}(1-p_{s})\sin^{2}\left(\zeta_{2}(t_{1},t_{2})\right)$.
This approximate expression is plotted in Fig.~\ref{fig:02} where
we observe a reasonable agreement for the position of the maxima and
minima of the SP.

\begin{figure}
\includegraphics[width=0.4\textwidth]{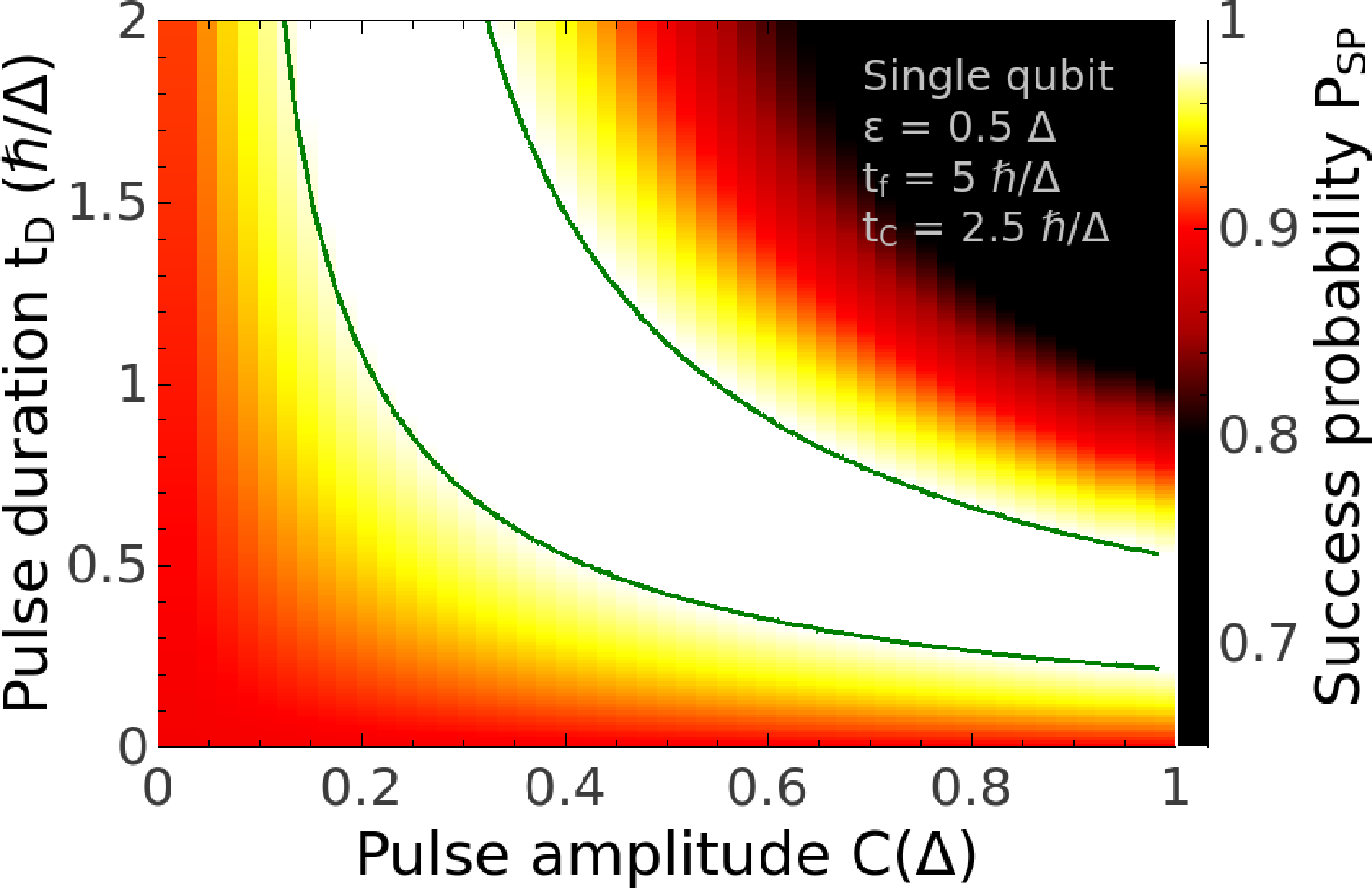}
\caption{(Color online) Contour plot of the success probability (SP), $P_{SP}$,
of a single qubit for fixed annealing time, $t_{f}=5\hbar/\Delta$,
and pulse center, $t_{C}=2.5\,\hbar/\Delta$, for varying the pulse
amplitude, $C$, and the pulse duration, $t_{D}$. The energy difference
between the ground and the excited states is $\varepsilon=0.5\,\Delta$,
at $C=0$. The green line encloses the area where $P_{SP}>0.98$.
In the case of $C=0$, the adiabatic condition does not hold.\label{fig:03}}
\end{figure}

In Fig.~\ref{fig:02} we consider the case where the SP for the
conventional QA is close to $1$ , implying that the adiabatic
condition holds for $C=0$. So now we focus on the case where
we decrease the annealing time, $t_{f}=5\hbar/\Delta$, and  
$\varepsilon=0.5\Delta$, then the SP for the conventional QA
 is decreased to $0.89$, due to the Landau-Zener transition. 
The minimum energy gap is $E_{G}^{min}=0.22\,\Delta$
at $t_{min}=4\,\hbar/\Delta$. Using Eq.~\eqref{eq:04} we find that
the right hand side has the value $3.6\hbar/\Delta$ which is close
to $t_{f}$, thus the adiabatic condition breaks. In Fig.~\ref{fig:03} 
we present a contour plot of the SP, $P_{SP}$, for varying the values of the pulse amplitude, $C$, and
the pulse duration, $t_{D}$, by keeping fixed the value of pulse
center, $t_{C}=2.5\,\hbar/\Delta$. We choose the pulse center value,
$t_{C}$, so as to be at the center of the annealing time. The SP,
$P_{SP}$, is increased, compared to its value for the conventional
QA, up to values of $1$. Furthermore, we observe that this increase
is persisted for a wide range of the values of the pulse parameters implying the robustness
of the proposed protocol. We also observe that for wider pulse durations,
$t_{D}$, we need smaller values of the pulse amplitude in order to
have an enhancement of the SP, compared to the $C=0$ case. The position
of the maxima of the SP are given by considering the maxima of the
$P_{SP,appr}$, which are connected with the points at which $\zeta_{2}(t_{1},t_{2})=N\pi$,
for $N=0,1,2\ldots$. The full expression for the $\zeta_{2}(t_{1},t_{2})$,
which can be analytically extracted from Eq.~\eqref{eq:06}, has
a complicate form and does not add anything to the discussion. In
Fig.~\ref{fig:03} we focus on the $N=0$ case for later reference.
The physical reason behind such result is the constructive
interference effect due to the pulse application, which effect is used
for increasing the SP for the multiqubit case as we see in the next
section.

\section{Increasing the success probability for multiqubit systems using pulsed quantum annealing\label{sec:III}}

\subsection{Multi-qubit model}

We now expand the discussion for the single qubit case to the multiqubit case to investigate the pulsed quantum annealing (PQA) protocol.  In the previous section we showed that changing the pulse parameters we can tune the SP for the case of a single qubit motivating the investigation of increasing the SP for the case of multiple qubits.

The Hamiltonian describing the PQA process for the multiqubit case is:
\begin{equation}
H=\frac{t}{t_{f}}H_{t}+C\Lambda_{P}(t)\sum_{i=1}^{n}\sigma_{z}^{i}+\left(1-\frac{t}{t_{f}}\right)\Delta\sum_{i=1}^{n}\sigma_{x}^{i},\label{eq:09}
\end{equation}
the third term represents the quantum fluctuation part, with an amplitude
$\Delta$, the second term is the diagonally applied diabatic pulse,
where the $\Lambda_{P}(t)$ scheduling is introduced in Eq.~\eqref{eq:01},
with a strength $C$, and $n$ is the number of qubits involved. The
quantum fluctuation $\sigma_{x}$ part gives a superposition state
in the computational basis ($\left|0\right\rangle $ and $\left|1\right\rangle $),
thus helps on exploring the energy landscape in order to find the
ground state. The first term is the  problem (spin-glass)
Hamiltonian: 
\begin{equation}
H_{t}=\sum_{i=1}^{n}\varepsilon_{i}\sigma_{z}^{i}+\sum_{i,j=1}^{n}J_{ij}\sigma_{z}^{i}\sigma_{z}^{j},\label{eq:10}
\end{equation}
where $\varepsilon_{i}$ and $J_{ij}$ are independent Gaussian random
numbers with zero mean and variance $\left\langle J_{ij}^{2}\right\rangle /\Delta^{2}=\left\langle \varepsilon_{i}^{2}\right\rangle /\Delta^{2}=1$.

Finding the ground state of $H_{t}$ is connected with minimizing
a cost function for an encoded optimization problem.
However, in reality, long annealing times cause unwanted decoherence
and dissipation effects, while short ones also induce nonadiabatic
Landau-Zener transitions which reduce SP. Applying a diabatic pulse,
like in the single qubit case, we can modulate the SP, of a multi-qubit
system and, for specific pulse parameters, enhance it, for a fixed
annealing time. We proceed by solving numerically the time-depended
Schr\"{o}dinger equation Eq.~\eqref{eq:02} using the Hamiltonian Eq.~\eqref{eq:09}
and its instantaneous eigenstates to calculate the SP, Eq.~\eqref{eq:03}.

\subsection{Five qubit example with low success probability}

\begin{figure*}[t]
\includegraphics[width=0.4\textwidth]{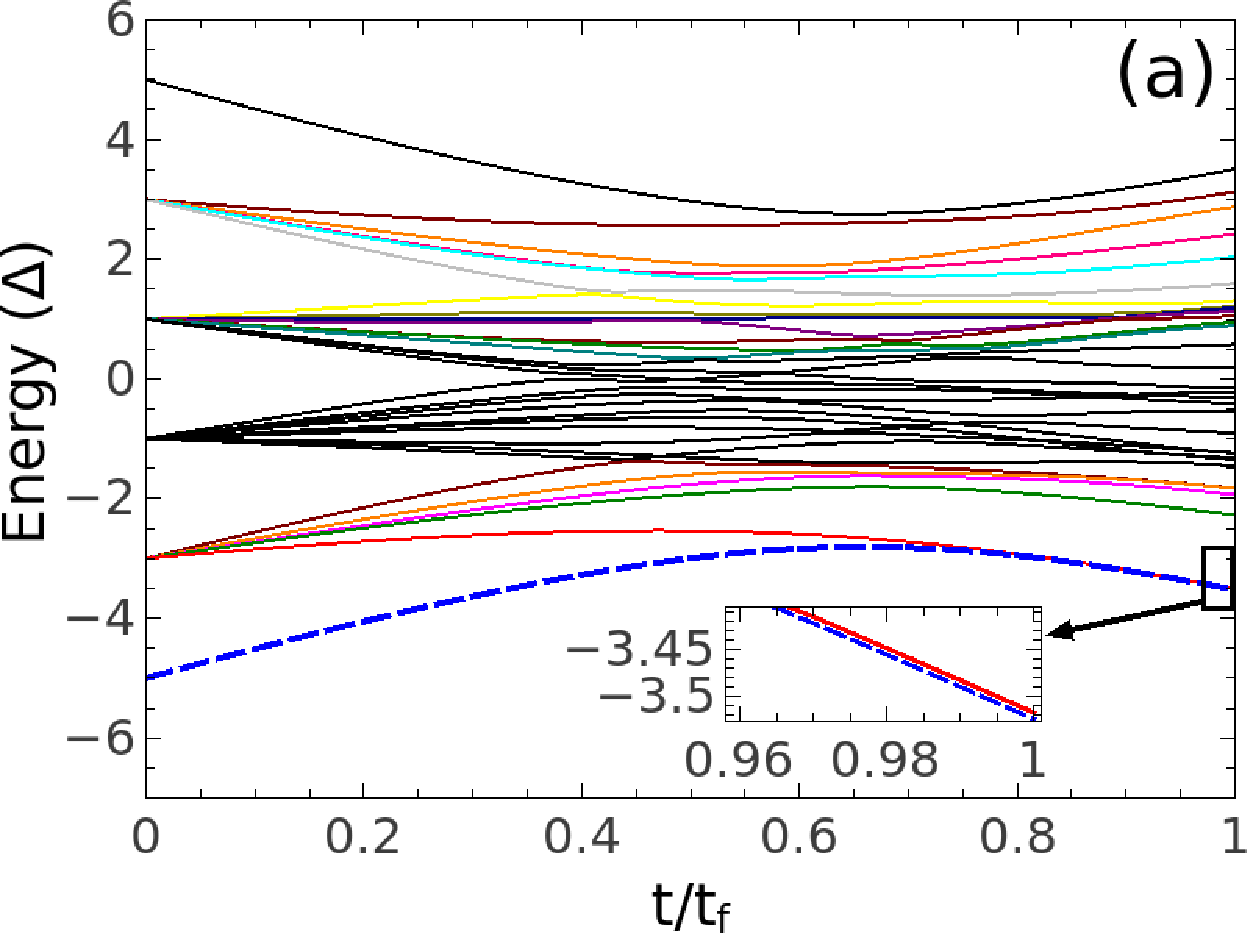}~~~\includegraphics[width=0.4\textwidth]{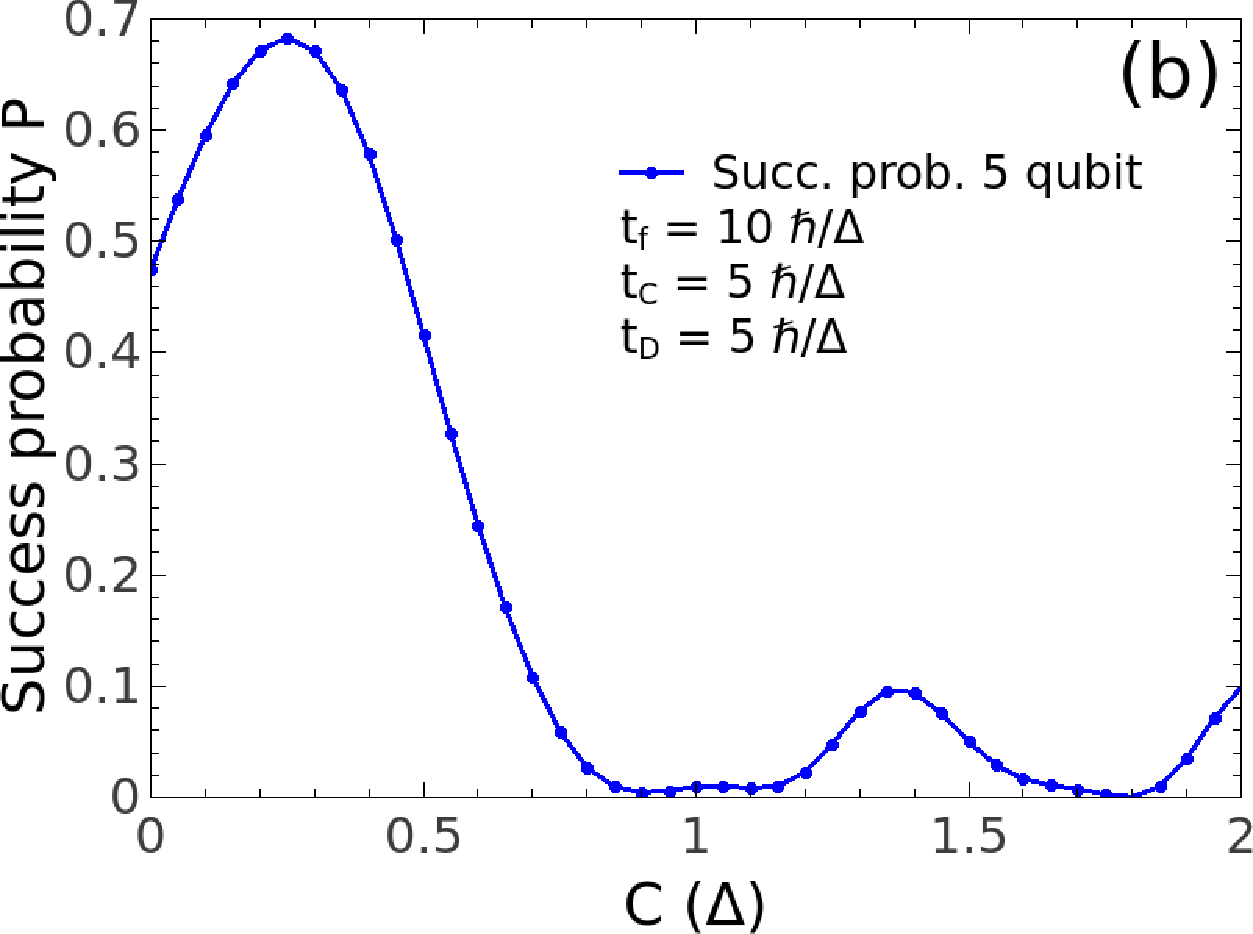}

\caption{(Color online) The success probability (SP) for the instance A
of 5 qubits and annealing time of $t_{f}=10\hbar/\Delta$. The SP for the 
conventional QA is $P_{S0}=0.47$, for $t_{f}=10\hbar/\Delta$. a) The
energy spectrum of the $2^{5}$ instantaneous eigenstates during the
annealing without the pulse application. Inset: Energy spectrum of the two lowest energy levels
close to the end of the annealing. b) The SP, $P$, for varying
the diabatic pulse amplitude, $C$, with fixed pulse center $t_{C}=5\hbar/\Delta$
and duration $t_{D}=5\hbar/\Delta$.\label{fig:04}}

\end{figure*}

We start by studying the SP of a $5$ qubit spin-glass instance, where
for the conventional QA the SP is $P_{S0}=0.47$, for an annealing
time of $t_{f}=10\hbar/\Delta$, for this instance the adiabatic condition
does not hold. We name this instance as instance A (iA). In Fig.~\ref{fig:04}(a)
we present the energy spectrum of iA for varying time $t$, we observe
that the ground state, thick blue dashed line, is very close to the
first excited state towards the end of the QA process,
check the inset of Fig.~\ref{fig:04}(a), 
thus Landau-Zener transitions induce a reduction of the SP. In order to increase the
SP we consider the effect of a single diabatic pulse application during
the QA to the $5$ qubit iA. In Fig.~\ref{fig:04}(b) the
pulse parameters are $t_{C}=5\hbar/\Delta$ and $t_{D}=5\hbar/\Delta$,
for the diabatic pulse center and duration respectively, while we
vary the pulse strength $C$. We observe that due to interference
effects, which are caused during the diabatic pulse application, the
SP oscillates for varying the pulse amplitude $C$. This effect can be connected
with the single qubit case, presented in Sec.~\ref{sec:II},
where varying the pulse amplitude induces constructive and destructive interference effects. 
For small pulse amplitudes, compared to the energy scale defined by the transition
amplitude $\Delta$, we observe that the SP increases compared to
the conventional QA. Further increasing the value of the amplitude
of the pulse amplitude, $C$, causes transitions to the higher excited states, thus
reducing the SP at the end of the annealing time.

\subsection{Multiple five qubit instances}

\begin{figure*}[t]

\includegraphics[width=0.4\textwidth]{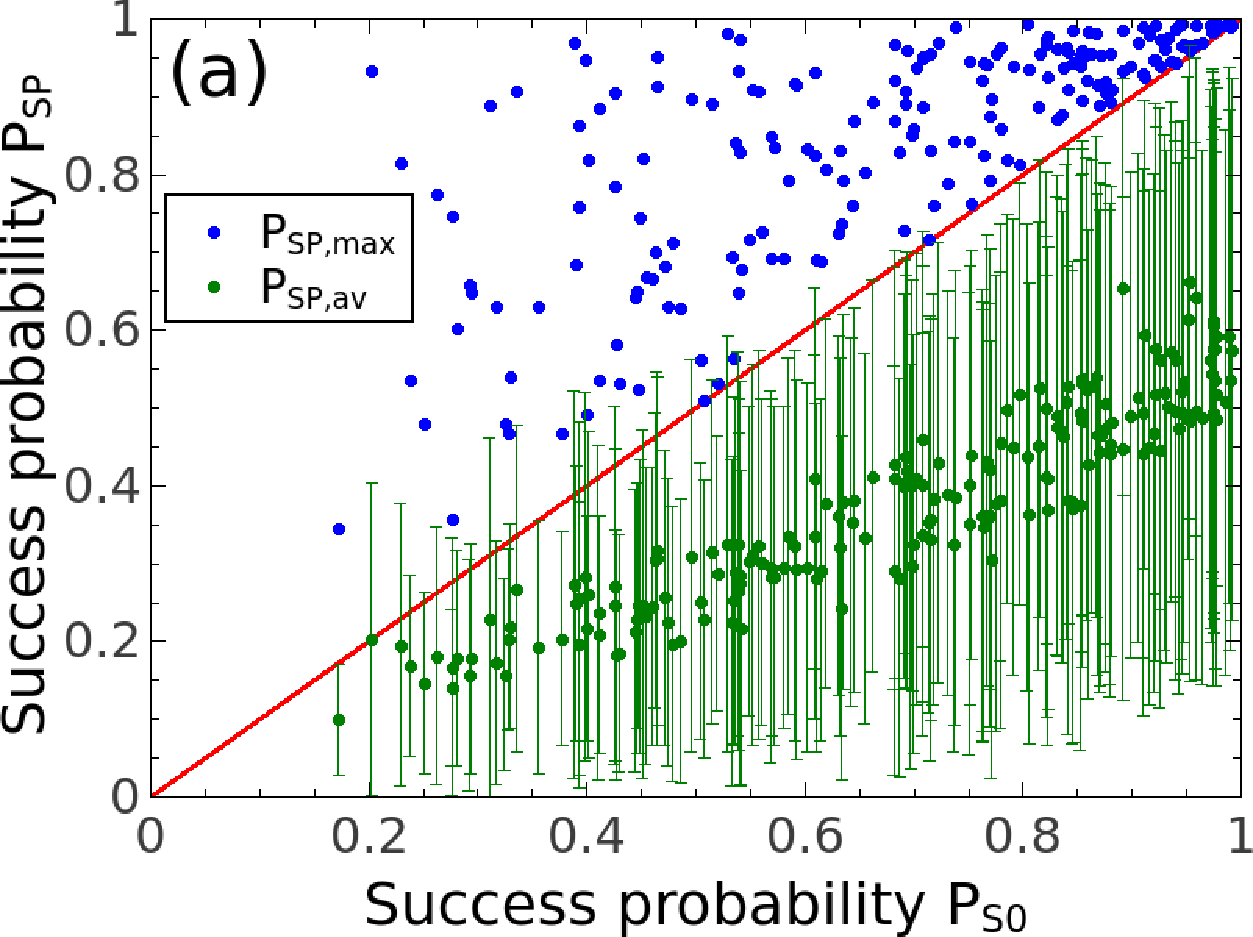}
\includegraphics[width=0.4\textwidth]{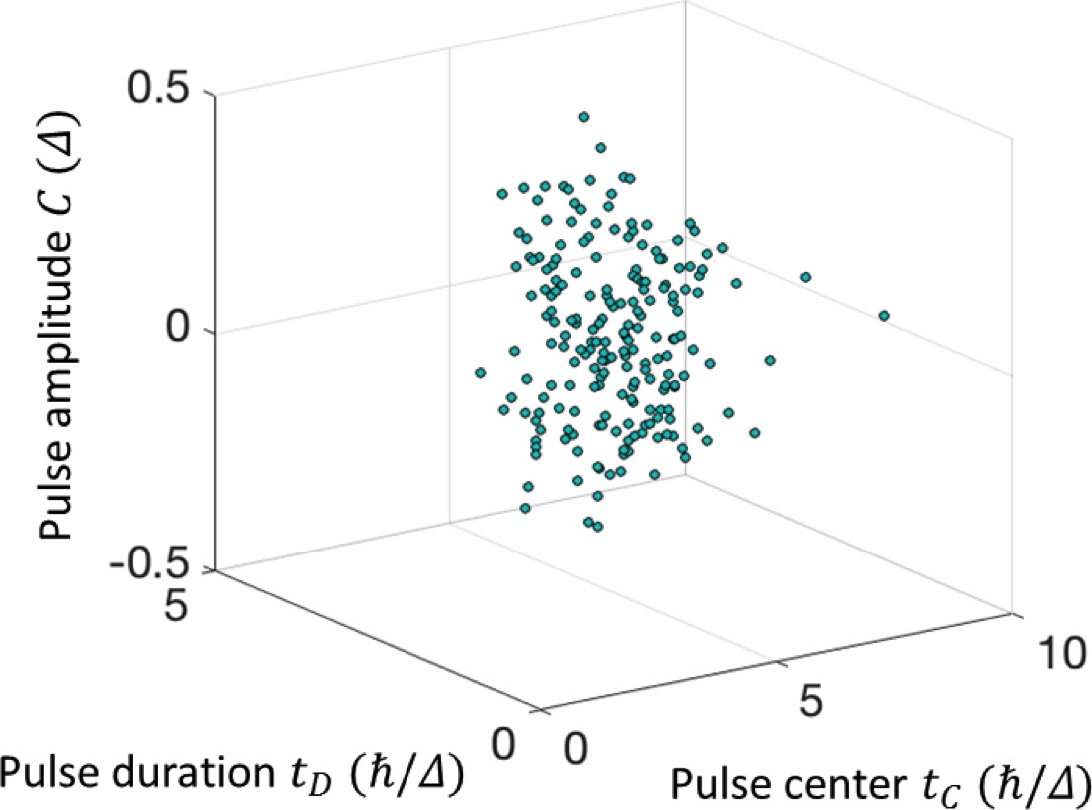}

\caption{(Color online) Plot of the success probability (SP) for $200$ instances
for the $5$ qubit case, where the annealing time is $t_{f}=10\hbar/\Delta$.
a) SP with the diabatic pulse, $P_{SP}$, compared without, $P_{S0}$.
Blue circles give the maximum attained SP, $P_{SP,max}$, while the
green give the averaged SP for the total sampling, $P_{SP,av}$. The
standard deviation is included as error bar in the figure. b) Averaged
values of the pulse parameters, pulse center $t_{C}$, pulse duration
$t_{D}$ and pulse amplitude $C$ for the parameter's value during
the sampling that $P_{SP}>P_{S0}$. \label{fig:05}}
\end{figure*}

From Fig.~\ref{fig:04} it has become clear that if we want to increase the SP 
for the PQA, compared to the conventional QA, we need to find the appropriate parameter set of the applied pulse
for each optimization problem. Let us define the SP when the
pulse is applied as $P_{SP}$ and without the pulse $P_{S0}$. In
Fig.~\ref{fig:05}(a) we present a plot of $P_{SP}$ to $P_{S0}$
for $200$ instances considering $5$ qubits, for fixed annealing
time $t_{f}=10\hbar/\Delta$. For each instance we run a routine for optimizing the parameters 
of the pulse, which samples $2706$ combinations 
for the pulse parameters for each instance, we define the maximum attained value of the SP as $P_{SP,max}$ and the averaged
SP over all samplings per instance as $P_{SP,av}$. With blue dots
we present the $P_{SP}^{max}$ versus $P_{S0}$ for each instance.
The thick red line shows the $P_{SP}=P_{S0}$ diagonal line and is
used as a guide for the reader, all instances that are above this
line present an increase in the SP for the PQA compared to
the conventional one. We observe that for all $200$ instances exist a set 
of pulse parameters that can increase the attained value of the SP, even 
for the cases that the initial SP is $P_{S0}>0.9$. The physical reason behind such a
remarkable behavior is connected with the constructive interference
during the diabatic pulse application, see Sec.~\ref{sec:II} detailed explanation of the relevant physics, for the specific set of parameters
for each instance. 

In Fig.~\ref{fig:05}(a) the green dots present $P_{SP,av}$
versus $P_{S0}$ for each instance, where we also show the standard
deviation over the sampling for each instance as error bars.
We observe that using the pulse application the average SP per instance $P_{SP,av}$
is reduced, compared to the conventional QA.

In Fig.~\ref{fig:05}(b) we present the average values of the pulse parameters for each 
instance that we have $P_{SP}>P_{S0}$, after the optimization routine, meaning that the
pulsed QA gives higher SP than the conventional QA for fixed annealing time. We observe
that the averaged pulse center has the value $t_{C}^{av}\sim t_{f}/2=5\hbar/\Delta$,
which means that the optimum pulse center $t_{C}$ is close to the
middle of the annealing, from the single qubit case we can observe, Fig.~\ref{fig:01}(a),
that the the linear ramping of the conventional annealing leads at an avoided energy level
crossing at around, and after, the $t/t_{f}\gtrsim0.5$. Thus, applying a pulse where
the avoided crossing exists is beneficial for creating a constructive
interference.

Furthermore, we notice that the optimum
amplitude parameter, $\left|C^{av}\right|$, is smaller than the energy
scale defined by the tunneling amplitude $\Delta$. Hence, the pulse
amplitude modulation enhances the SP and can have a peak for small
$C$ value, compared to $\Delta$. Moreover, we observe that half of the generated instances
are enhanced for the case we have positive pulse strength, $C$, while
the other half of the instances for negative. Therefore, the enhancement
of the SP is not due to the energy gap opening but due to the constructive
interference during the pulse application. Based on the above remarks
we can reduce the sampling parameter space considerably, considering
mainly pulses of small amplitude $C$ with a pulse center close to
$t_{f}/2$, although we make a broader sampling for the pulse center.
In Fig.~\ref{fig:05} the sampling space is $O(10^{3})$ after following
the above remarks the sampling space can be reduced to $O(10^{1})$.

\begin{figure*}[t]
\includegraphics[width=0.4\textwidth]{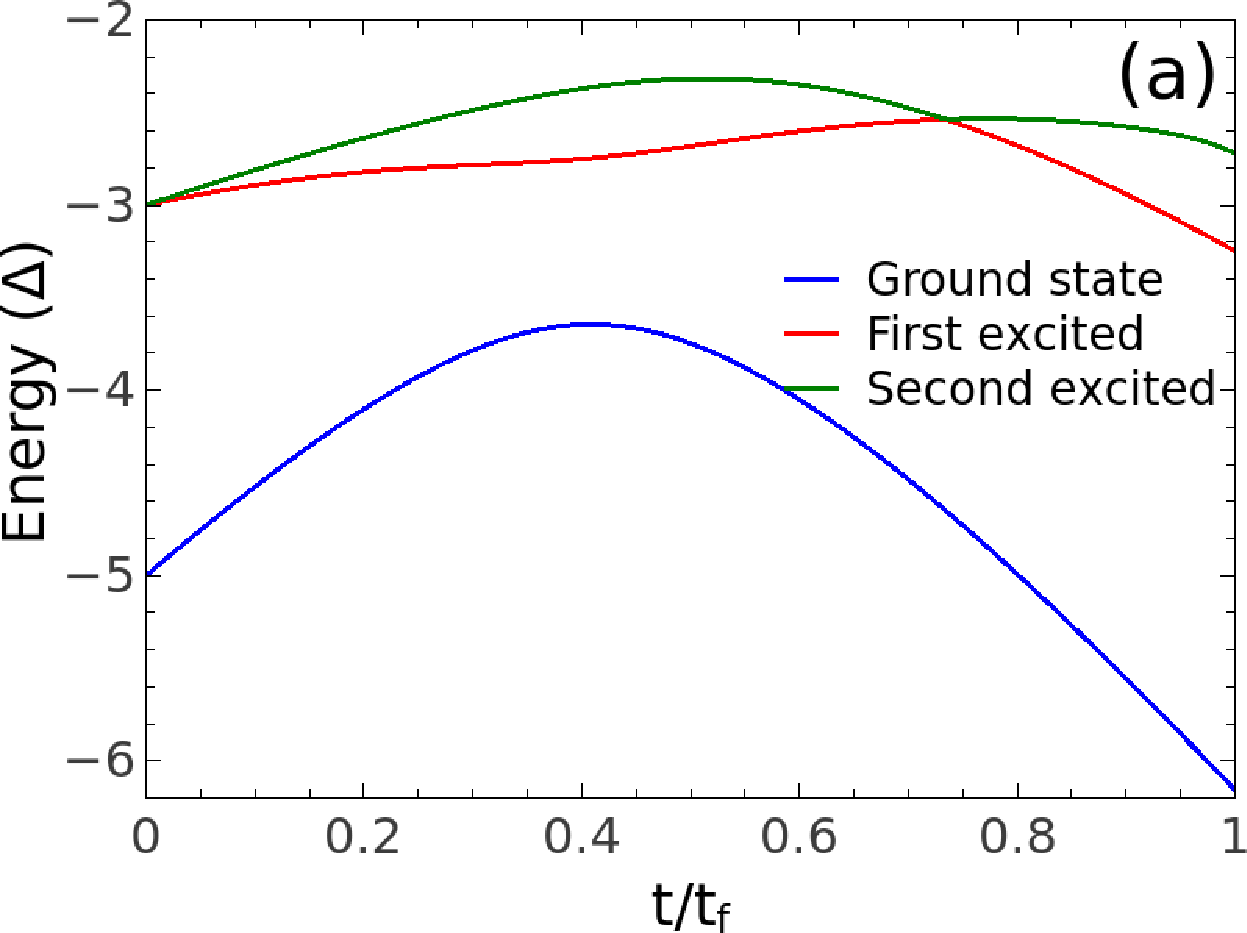}~~~\includegraphics[width=0.4\textwidth]{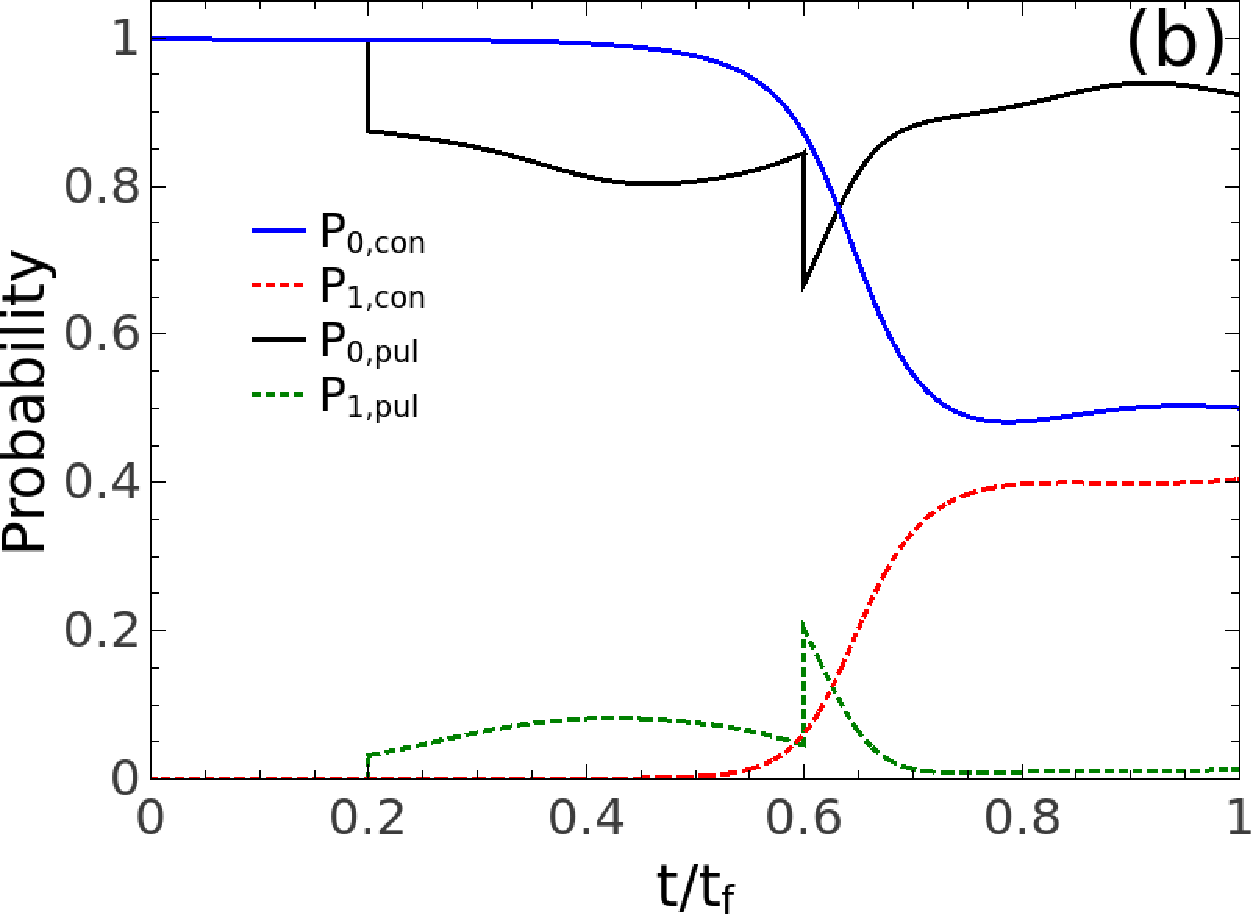}

\caption{(Color online) (a) The three lowest energy states and (b) the
projection probability of the ground and excited instantaneous eigenstates
to the system state $P_{i}(t)=\left|\left\langle \psi_{i}(t)\left|\psi(t)\right.\right\rangle \right|^{2}$,
for $i=0,1$ respectively. We consider the iB $5$ qubit instance. $P_{i,con}(t)$ and $P_{i,pul}(t)$
are the state probability overlap for the conventional and pulsed
QA, respectively. The annealing time is $t_{f}=10\hbar/\Delta$.
\label{fig:06}}
\end{figure*}

In order to further understand the behavior of the PQA protocol 
we present in Fig.~\ref{fig:06} a glassy instance for the 5 qubit system. 
After the optimization process, the parameters of the pulse enhancing the SP are: $t_{C}=4\hbar/\Delta$,
$t_{D}=4\hbar/\Delta$ and $C=0.3\Delta$. We name this instance as instance B (iB). 
In Fig.~\ref{fig:06}(a) we present the three lowest energy states of the iB, and
in Fig.~\ref{fig:06}(b) we present the probability
overlap of the ground, $i=0$, and first excited, $i=1$, instantaneous
eigenstates to the system state evolution, 
$P_{i}(t)=\left|\left\langle \psi_{i}(t)\left|\psi(t)\right.\right\rangle \right|^{2}$.

For the instance iB, we observe that there is a steep avoided crossing, Fig.~\ref{fig:06}(a),
around $t/t_{f}=0.4$, which cause a Landau-Zener transition that
reduces the SP at the end of the annealing. The optimized pulse center,
$t_{C}$, is at the avoided crossing. In Fig.~\ref{fig:06}(b) we observe that initially the system is at
the ground state, until approaching the avoided crossing which causes a state mixing, 
between the ground and first excited states. This finally leads to a reduction of SP at 
the end of the annealing. In direct analogy with the single qubit case, presented in Sec.~\ref{sec:II}, the pulse application causes a constructive interference during 
the pulse application, leading to an increase in the SP at the end of the annealing, comparing to the 
conventional QA protocol for fixed annealing time.

The next step is to reduce the annealing time, $t_{f}$, in order
to further reduce the SP for the conventional QA, thus allowing 
further space for enhancement for the PQA for showing clearly 
the importance of implementing the PQA protocol.

\begin{figure}
\includegraphics[width=0.45\textwidth]{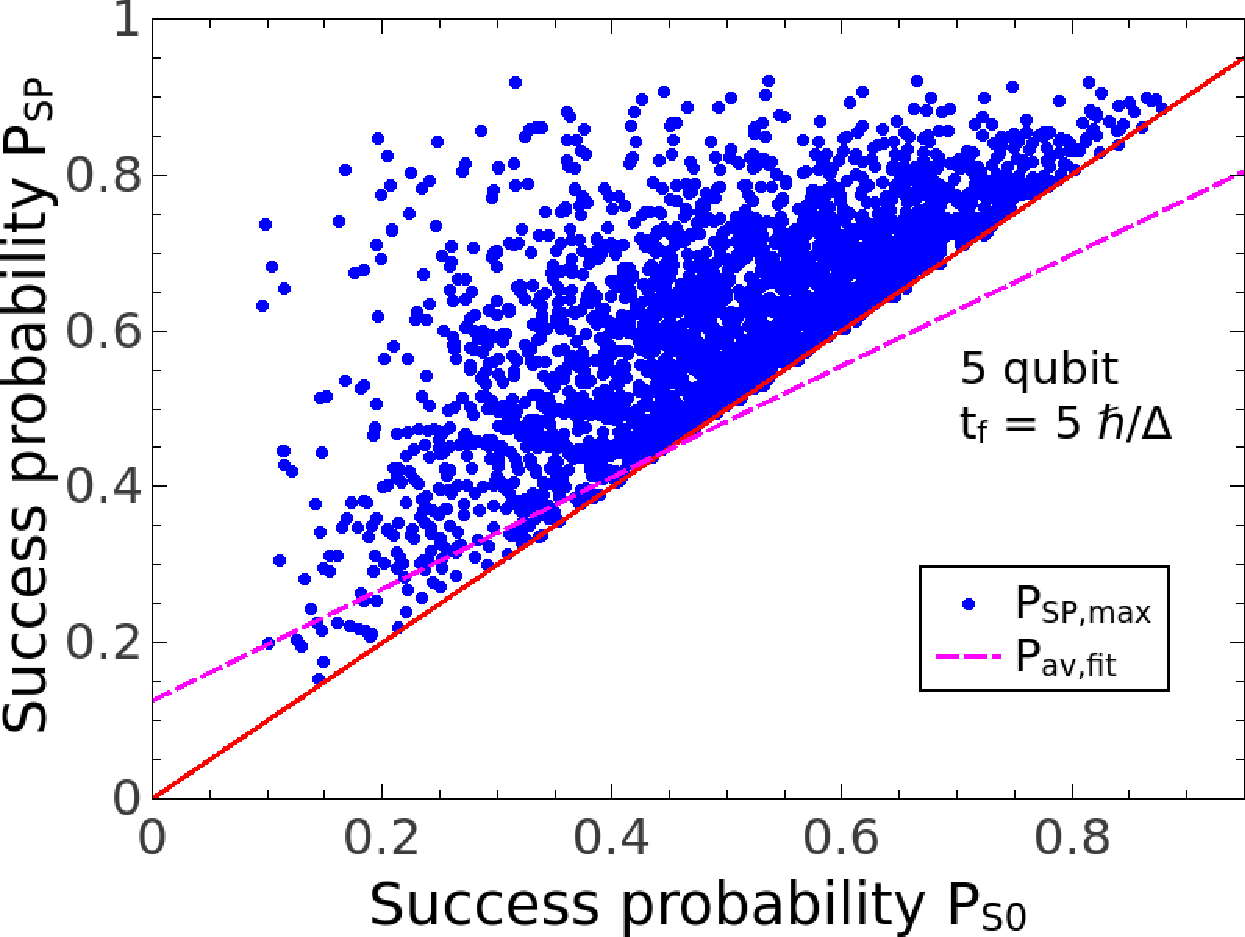}
\caption{(Color online) Plot of the success probability (SP) with the diabatic
pulse, $P_{SP}$, compared without the pulse, $P_{S0}$, QA for $1000$
instances for the $5$ qubit case, the annealing time is $t_{f}=5\hbar/\Delta$. 
The blue circles give the maximum attained
SP, $P_{SP,max}$, to the $P_{S0}$. The magenta dashed line is a linear fitting of
the $P_{SP,av}$ to the $P_{S0}$, $P_{SP,av}=A\cdot P_{S0}+B$, the
linear fitting parameters are $A=0.72$ and $B=0.13$. \label{fig:07}}
\end{figure}

In Fig.~\ref{fig:07} we present the SP for $1000$ instances
of $5$ qubit systems, with the diabatic pulse application, $P_{SP}$,
compared without the pulse, $P_{S0}$, 
for an annealing time of $t_{f}=5\hbar/\Delta$;
for this annealing time the adiabatic condition is violated for the conventional
QA. The blue circles give the maximum attained 
value of the SP, $P_{SP,max}$, where 
the sampling number for the pulse parameter optimization 
is just $90$. We observe that 
the instances that $P_{SP,max}>P_{S0}$ is $100\%$, where on the same time for
$80\%$ of the instances we have $(P_{SP,max} - P_{S0})/P_{S0}> 5\%$.
These results support that for fixed annealing times we can enhance
the SP for specific set of pulse parameters for each instance.

In Fig.~\ref{fig:07} we also present a linear fit, of the form $Ax+B$,
of the averaged SP over the sampling of the pulse parameters, per instance 
$P_{SP,av}$ considering as $x=P_{S0}$. The 
linear fit parameters are $A=0.72$ and $B=0.13$. Moreover, in Fig.~\ref{fig:07} we present the guide red line $P_{SP}=P_{S0}$
and we observe this line crosses the
fitting of  $P_{SP,av}$ vs $P_{S0}$. For the annealing time $t_{f}=5\hbar/\Delta$
considered the cross is at $P_{S0}^{c}=0.45$. 
For $P_{S0}^{c}>P_{S0}$ the averaged SP,
$P_{SP,av}$, is increased for the case of pulsed over the conventional
QA.

\subsection{Scaling of success probability for the multi-qubit case}

Up to now we focus on the case of $5$ qubits to analyze and understand
the influence of the diabatic pulse to the SP. It is desirable to investigate how the proposed diabatic PQA 
scales as the number of qubits $n$ increases. We consider
the cases of annealing time of $t_{f}=5\hbar/\Delta$, with a sampling
of the pulse parameters of $90$ samples for each instance,
where the the highest value for $P_{SP,max}$ per
instance is recorded. In order to check this behavior we introduce
the relative success probability per multiqubit system
\begin{equation}
R_{SP,max}=\frac{\bar{P}_{SP,max}-\bar{P}_{S0}}{\bar{P}_{S0}},\label{eq:11}
\end{equation}
where $\bar{P}$ is the averaged SP over all instances and the
sampling over the pulse parameters.
 $R_{SP,max}$ is a measure of how much the SP increases or
decreases for the PQA compared to the conventional QA.

\begin{figure*}[t]

\includegraphics[width=0.4\textwidth]{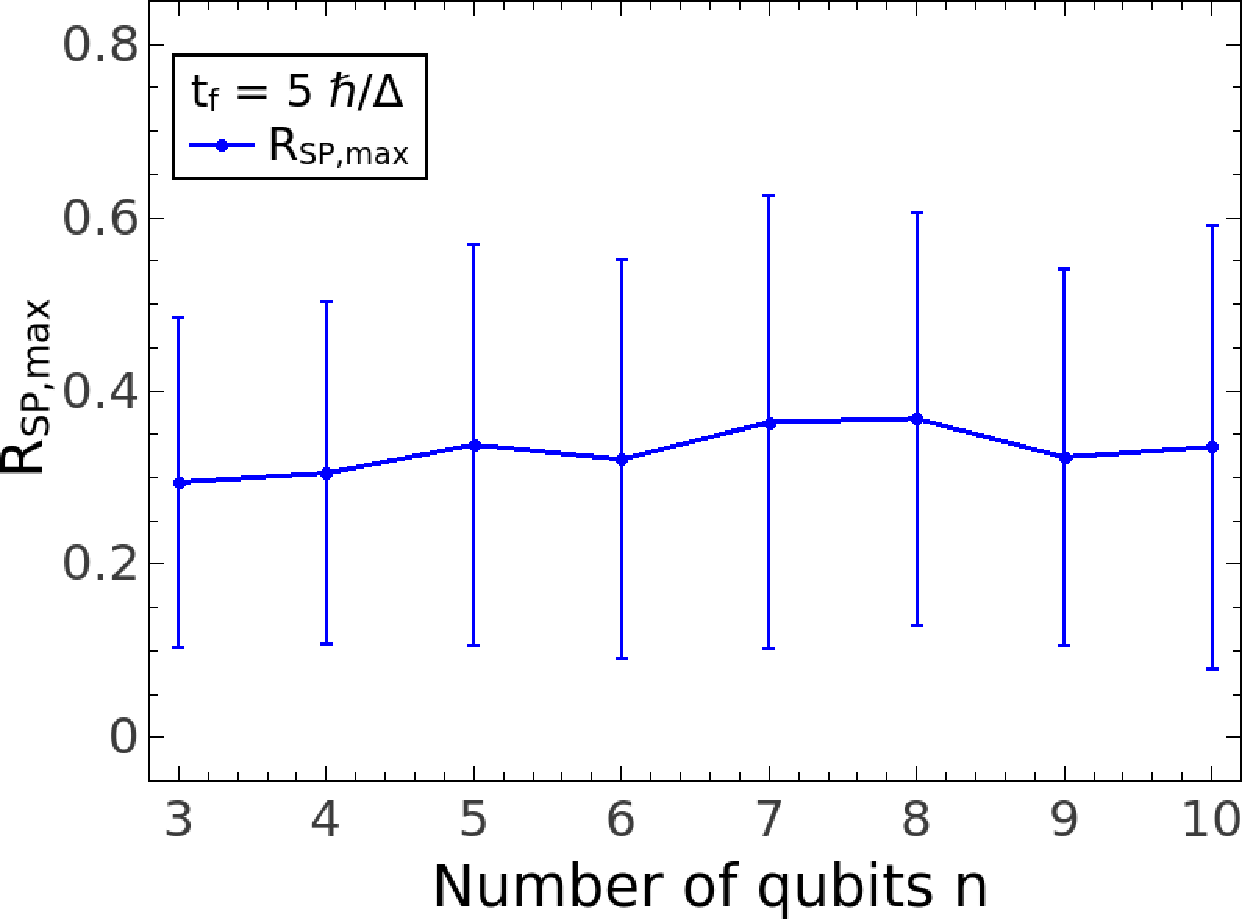}~~~\includegraphics[width=0.4\textwidth]{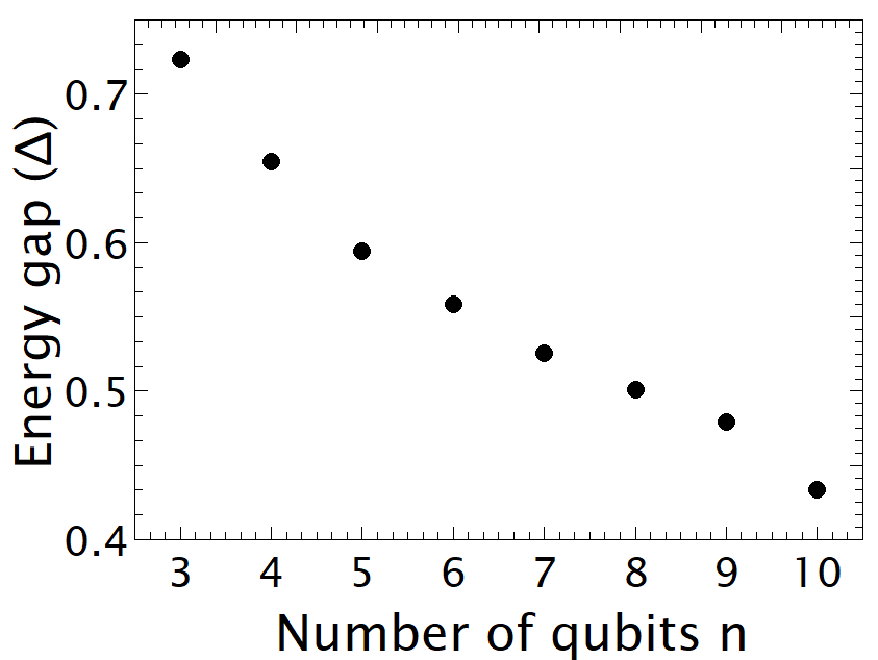}

\caption{(Color online) (a) Plot of the averaged relative success probability (SP),
$R_{SP,max}$, for varying the number of qubits $n$. The standard deviation is 
presented as error bars, which
is multiplied by $0.5$. The annealing time is $t_{f}=5\hbar/\Delta$. 
The numerical results are given as dots which are connected by lines in order to provide a guide to
the eye. (b) Minimum of the energy gap between the ground and first excited state for
varying the number of qubits $n$, each point is the average of 10000 random instances.\label{fig:08}}
\end{figure*}

In Figs.~\ref{fig:08}a we present the value of $R_{SP,max}$,
blue dots connected by blue lines, for increasing number of qubit
$n$, for annealing times $t_{f}=5\hbar/\Delta$.  
Instances of up to $n=10$ qubits are considered. We observe that for increasing $n$ 
the averaged $R_{SP,max}$ 
approaches a value of $30\%$.  In Fig.~\ref{fig:08}b we present the averaged 
minimum energy between the ground and excited states by varying the qubit number $n$, where 10000 instances per qubit number have been averaged.
 We observe that the energy gap, between the ground and the
first excited state, decreases as the number of qubits is increased,
thus making it harder for the conventional quantum annealing to find
the ground state, for the annealing time discussed in this section.
The fact that the averaged SP $R_{SP,max}$ is positive shows that the PQA 
outperforms the conventional one. For an annealing time of $t_{f}=5\hbar/\Delta$
the maximum relative SP, $R_{SP,max}$, shows a slight steady increase
as the qubit number $n$ increases. Of course the averaged relative
SP, $R_{SP,max}$, suffers from high standard
deviation, which is plotted as error bars in Fig.~\ref{fig:08}a,
multiplied by $0.5$.

The data presented in this paper are for up to $n=10$ qubit systems,
since we use a sampling to find the appropriate pulse
parameters that provide an enhancement of the SP. 
We expect the scaling
to sustain for increasing $n$, due to the fact that we
consider a closed system at zero temperature. This approximation is
justified for the short annealing times considered. 
In order to check the effectiveness of the PQA protocol for practical cases, e.g.,
D-Wave's quantum annealing machine, more detailed numerical study for large spin systems ($n>10$) is needed. 
Such a systematic numerical study is an interesting future problem.

Finally it should be mentioned that the diabatic
pulse introduced in the diagonal part of the Hamiltonian 
can be implemented in D-Wave superconducting quantum annealing machines based on a superconducting flux qubit.

\section{Summary and Discussion\label{sec:IV}}

In this paper we start by investigating the success probability (SP)
of a single qubit for the case of the conventional quantum annealing
(QA) plus a diabatic pulse application. The diabatic pulse application
can modulate the SP, by varying the pulse parameters, and enhance
it for specific set of parameters. The constructive and destructive
interference effect due to the acquired phase during the pulse application
is the physical reason for such behavior of the system. Using the
transfer matrix method, combined with the sudden approximation, we
were able to present such an effect in a semi-analytical manner.

For the multiqubit cases, the pulse application can modulate the SP and
for specific set of parameters we can increase the SP, compared to the conventional
QA. When a number of random (spin-glass) $5$ qubit instances is considered,
for each one of them we were able to increase the SP, for specific
set of pulse parameters. Furthermore, reducing the annealing time,
$t_{f}$, and based on the physical observations, that the avoided
crossing emerges at times around and later from the middle of the annealing process,
and consider pulse amplitudes smaller than the quantum fluctuations, 
we are able to reduce the sampling space.

As the number of qubits, $n$, is increased we have an enhancement
of the SP. This enhancement seems to be constant although an improvement
above $30\%$ is persistent as the number of qubits is increased. 
This behavior is justified by the fact that increasing
$n$ the energy gap between the ground and excited
states decreases, thus reducing the SP for the conventional QA, and
giving space for larger enhancement for the PQA. This effect is studied
by introducing the relative SP, $R_{SP}$, and investigate how this
quantity behaves with increasing $n$. The averaged maximum attained
SP, per sampling over all instances, per number of qubits shows better
performance for the case of pulsed over the conventional QA after
optimizing the pulse parameters.

The main message of our results is that
breaking away from adiabaticity we are able to present an increase
in the SP in shorter annealing times by applying a pulse. We have
two sources for this increment. Reduction of the annealing time reduces
the interaction of the system with the environment, thus reducing
decoherence and dissipation effects. On the same time, reducing the
annealing time significantly we can run the appropriate number of
samplings, while increasing the SP, within a reasonable time frame.

Recently, D-Wave Systems Inc. announced the implementation of the
reverse QA protocol \cite{D-Wave2017}, where the initial state is
not the usual ground state of the quantum fluctuations, but a specific
state which might be closer to the ground state of the problem Hamiltonian.
In Ref.\cite{Chancellor2017} it is proposed a hybrid computing method
where initially a classical algorithm, like simulated annealing, can
be used to find a solution close to the real ground state of a complicated
problem. Then, this state can be used as the initial state for applying
the reverse QA protocol for a local search in the phase space. The
PQA protocol can also be used in a similar manner, to provide
if not the real ground state, a state very close to it. This is a
different research path to follow in the future for a PQA machine
application.

One year after submitting our manuscript in arXiv:1806.08517 (submission date 22 Jun. 2018) we became aware of ref. \cite{Munoz-Bauza2019} (submission date 18 March 2019)
where PQA for only the single qubit case was investigated using the interferometer-interpretation we gave in Fig. 1(b).

\begin{acknowledgment}

The authors would like to thank K. Imafuku, M. Maezawa, Y. Seki, and S. Tanaka
for useful discussions. This work is based on results obtained from
a project commissioned by the New Energy and Industrial Technology
Development Organization (NEDO), Japan.

\end{acknowledgment}

\appendix
\section{Sudden approximation\label{sec:Appendix-A}}
Following the refs.\cite{Messiah2014,Silveri2015} lets consider the
case where we have the diabatic pulse application at the time $t_{1}$,
for times $t<t_{1}$ the Hamiltonian has the form
\begin{equation}
\begin{aligned}
H_{0}(t){} & =\left(\begin{array}{cc}
t/t_{f}\varepsilon & \left(1-t/t_{f}\right)\Delta\\
\left(1-t/t_{f}\right)\Delta & -t/t_{f}\varepsilon
\end{array}\right) \\
& =E_{G}^{0}(t)\left(\begin{array}{cc}
\cos\left(\theta_{0}(t)\right) & \sin\left(\theta_{0}(t)\right)\\
\sin\left(\theta_{0}(t)\right) & -\cos\left(\theta_{0}(t)\right)
\end{array}\right),
\label{eq:A01}
\end{aligned}
\end{equation}
where we define $\cos\left(\theta_{0}(t)\right)=\left(t/t_{f}\right)\varepsilon/E_{G}^{0}(t)$
and $\sin\left(\theta_{0}(t)\right)=\left(1-t/t_{f}\right)\Delta/E_{G}^{0}(t)$,
$E_{G}^{0}(t)=\sqrt{\left(t/t_{f}\right)^{2}\varepsilon^{2}+\left(1-t/t_{f}\right)^{2}\Delta^{2}}$
is the energy gap between the ground and the excited states. The relevant
eigenstates are 
\begin{equation}
\left|\psi_{0}^{-}(t)\right\rangle =\left(\begin{array}{c}
-\sin\left(\theta_{0}(t)/2\right)\\
\cos\left(\theta_{0}(t)/2\right)
\end{array}\right),\:\:\left|\psi_{0}^{+}(t)\right\rangle =\left(\begin{array}{c}
\cos\left(\theta_{0}(t)/2\right)\\
\sin\left(\theta_{0}(t)/2\right)
\end{array}\right),\label{eq:A02}
\end{equation}
where $\left|\psi_{0}^{-}(t)\right\rangle $ is the ground and $\left|\psi_{0}^{+}(t)\right\rangle $
the excited states. Similarly for $t>t_{1}$ we have the Hamiltonian
\begin{equation}
\begin{aligned}
H_{C}(t) {} & = \left(\begin{array}{cc}
t/t_{f}\varepsilon+C & \left(1-t/t_{f}\right)\Delta\\
\left(1-t/t_{f}\right)\Delta & -t/t_{f}\varepsilon-C
\end{array}\right) \\
    & = E_{G}^{C}(t)\left(\begin{array}{cc}
\cos\left(\theta_{C}(t)\right) & \sin\left(\theta_{C}(t)\right)\\
\sin\left(\theta_{C}(t)\right) & -\cos\left(\theta_{C}(t)\right)
\end{array}\right),\label{eq:A03}
\end{aligned}
\end{equation}
where we defined $\cos\left(\theta_{C}(t)\right)=\left(t/t_{f}\varepsilon+C\right)/E_{G}^{C}(t)$
and $\sin\left(\theta_{C}(t)\right)=\left(1-t/t_{f}\right)\Delta/E_{G}^{C}(t)$,
$E_{G}^{C}(t)=\sqrt{\left(t/t_{f}\varepsilon+C\right)^{2}+\left(1-t/t_{f}\right)^{2}\Delta^{2}}$
is the energy gap between the ground and the excited states in the
presence of the pulse of amplitude $C$. The relevant eigenvectors
are
\begin{equation}
\left|\psi_{C}^{-}(t)\right\rangle =\left(\begin{array}{c}
-\sin\left(\theta_{C}(t)/2\right)\\
\cos\left(\theta_{C}(t)/2\right)
\end{array}\right),\:\:\left|\psi_{C}^{+}(t)\right\rangle =\left(\begin{array}{c}
\cos\left(\theta_{C}(t)/2\right)\\
\sin\left(\theta_{C}(t)/2\right)
\end{array}\right),\label{eq:A04}
\end{equation}
where $\left|\psi_{C}^{-}(t)\right\rangle $ is the ground and $\left|\psi_{C}^{+}(t)\right\rangle $
the excited states.

At the time $t=t_{1}$ due to the sudden transition of the Hamiltonian
from the form $H_{0}(t)$ to $H_{C}(t)$ the system stays in the same
state, then the matrix $N_{1}$ can be used to describe the transition
from the basis $\left|\psi_{0}^{\pm}(t=t_{1})\right\rangle $ to the
basis $\left|\psi_{C}^{\pm}(t=t_{1})\right\rangle $, thus the relevant
state mixing between the ground and the excited states. The elements
of the matrix $N_{1}$ are $\sqrt{p_{s}}=\sin\left(\theta_{C}(t_{1})-\theta_{0}(t_{1})\right)$
and $\sqrt{1-p_{s}}=\cos\left(\theta_{C}(t_{1})-\theta_{0}(t_{1})\right)$
and the system state, $\mathbf{b}(t)=\left(\begin{array}{c}
b_{0}(t)\\
b_{1}(t)
\end{array}\right)$, at $t_{1}^{+}$ is given by
\begin{equation}
\left(\begin{array}{c}
b_{0}(t_{1}^{+})\\
b_{1}(t_{1}^{+})
\end{array}\right)=\left(\begin{array}{cc}
\sqrt{1-p_{s}} & \sqrt{p_{s}}\\
\sqrt{p_{s}} & \sqrt{1-p_{s}}
\end{array}\right)\left(\begin{array}{c}
b_{0}(t_{1}^{-})\\
b_{1}(t_{1}^{-})
\end{array}\right).\label{eq:A05}
\end{equation}
Most of the readers would anticipate the physical evolution of the
system under the QA protocol to be determined by the Landau-Zener
transitions. Then, at the diabatic pulse application times the faulty
impression is that we would had a transition to the excited state
with probability $1$. This is wrong as we can see from the full numerical
data and the explanation given from the transfer matrix approach,
using the sudden approximation. The pulse application discussed in
this paper is a diabatic process thus the effect of the Landau-Zener
physics is irrelevant at $t_{1}$ and $t_{2}$.

\end{document}